\documentclass[a4paper]{jpconf}
\usepackage[usenames,dvipsnames]{color}
\usepackage{amsmath}
\usepackage{soul,ragged2e}
\usepackage{graphicx, caption, subcaption}
\usepackage{hyperref}
\hypersetup{
	unicode=true,
	pdfnewwindow=true,
	colorlinks=true, 	% (false,true)
	pdfborder={0 0 0},
	linkcolor=blue,
	linktoc=all, 		% (none,all)
	citecolor=blue,
	urlcolor=blue,
	breaklinks=true
}

\graphicspath{{"img/"}}
\newcommand*{\thead}[1]{\multicolumn{1}{c}{#1}}

\begin{document}

\title{Active flow control of a turbulent separation bubble through deep reinforcement learning}

\author{Bernat Font$^{1,2,*}$, Francisco Alcántara-Ávila$^3$, Jean Rabault$^4$, Ricardo Vinuesa$^3$ and Oriol Lehmkuhl$^1$}

\address{$^1$ Computer Applications in Science and Engineering (CASE), Barcelona Supercomputing Center, Barcelona, Spain}
\address{$^2$ Faculty of Mechanical Engineering, TU Delft, Delft, The Netherlands}
\address{$^3$ FLOW, Engineering Mechanics, KTH Royal Institute of Technology, Stockholm, Sweden}
\address{$^4$ Independent researcher, Oslo, Norway}

\ead{b.font@tudelft.nl}

\begin{abstract}
The control efficacy of classical periodic forcing and deep reinforcement learning (DRL) is assessed for a turbulent separation bubble (TSB) at $Re_\tau=180$ on the upstream region before separation occurs. The TSB can resemble a separation phenomenon naturally arising in wings, and a successful reduction of the TSB can have practical implications in the reduction of the aviation carbon footprint. We find that the classical zero-net-mas-flux (ZNMF) periodic control is able to reduce the TSB by 15.7\%. On the other hand, the DRL-based control achieves 25.3\% reduction and provides a smoother control strategy while also being ZNMF. To the best of our knowledge, the current test case is the highest Reynolds-number flow that has been successfully controlled using DRL to this date. In future work, these results will be scaled to well-resolved large-eddy simulation grids. Furthermore, we provide details of our open-source CFD--DRL framework suited for the next generation of exascale computing machines.
\end{abstract}

\section{Introduction}
\vspace*{0.25cm}
Wall turbulence, a phenomenon naturally arising in the field of fluid dynamics, has long captivated the curiosity of scientists and engineers alike. Although our knowledge about turbulence has markedly advanced over the past decades, there are still numerous open questions that continue to challenge researchers in the field. In the area of flow control (FC), a better understanding of turbulence is the key to unlock novel control strategies that can help us optimize different processes or systems. FC plays a crucial role in the transportation industry, potentially allowing the reduction of the carbon footprint by improving the efficiency of transport vehicles. By gaining insights into the underlying mechanisms of wall turbulence and its impact on aerodynamic drag and skin friction, FC strategies can be derived and applied in a passive or active manner. The efficacy of this approach lies in the fact that even a minor reduction in drag can result in a substantial decrease in the overall percentage of global emissions. A practical situation in which FC can help reduce CO\textsubscript{2} emissions arises in the aeronautical sector. Consider the upper part of an aircraft wing, also known as the suction side, which exhibits an adverse pressure gradient (APG) that can induce flow separation at high angles of attack. A turbulent separation bubble (TSB) is formed when the flow is separated and then reattached. This situation may occur mostly during take-off and landing on the high-lift devices (flaps, slats), or during cruise when the incoming free-stream flow is highly perturbed. In order to alleviate constraints on the aircraft design and facilitate a more precise optimization of the cruise phase, flow control can potentially help to reduce the TSB hence improving flow attachment. There are countless applications of FC as referenced in the literature, and the reader is referred to Refs. \cite{gad91,cat11} for comprehensive reviews.

A turbulent boundary layer (TBL) subjected to an APG (APG TLB) that generates a TSB resembles the suction side of a wing in a simplified manner, and it is the most suitable first approach to learn the key features to control separation. APG TBLs are characterized by a wider parametric space compared to zero-pressure gradient (ZPG) TBLs. The most influential parameters are \cite{mon11}: a) the Rotta--Clauser pressure-gradient parameter $\beta=(\delta^*/\tau_w)\mathrm{d}_x P$, where $\tau_w$ is the wall-shear stress, $P$ is the static pressure, $x$ is the streamwise direction coordinate, and $\delta^*$ is the displacement thickness -- this parameter represents the pressure-to-viscous force ratio. b) the friction Reynolds number $Re_\tau=\delta^* u_\tau/\nu$, where $u_\tau=\sqrt{\tau_w / \rho}$ is the friction velocity, $\rho$ is the fluid density, and $\nu$ is the kinematic viscosity -- this parameter represents the large-scale to small-scale ratio of wall turbulence. c) the acceleration parameter $K=(\nu/u_\infty^2)\mathrm{d}_x u_\infty$, where $u_\infty$ is the local free-stream velocity -- this parameter represents the equilibrium state of the boundary layer. A number of examples in the literature have investigated APG TBLs under different parametric regimes, both numerically and experimentally. Some examples include \cite{kit16, abe17, kit17, vin17, wu_20} using direct numerical simulation (DNS), \cite{bob16, bob17, poz22} using well-resolved large-eddy simulation (LES), and \cite{mon11, har13, san20} using experiments.

The application of active flow control (AFC) for a TSB is typically performed through surface actuators located upstream of the separation bubble. AFC has been the focus of numerous TSB studies for both canonical and practical flows. The latter refer to wing-like configurations, where the surface curvature and/or angle of attack induce flow separation. You and Moin \cite{you08} applied oscillatory synthetic jets, also known as zero-net-mass-flux periodic control (or periodic forcing), which connected the pressure and suction sides of a NACA 0015 airfoil. By harmonic blowing and suction, they found that this control effectively delayed the separation of the boundary layer at high angles of attack. Atzori {\it et al.} \cite{atz21} focused on uniform blowing and uniform suction applied to the suction side of a NACA 4412 airfoil, finding that AFC had a larger impact on the APG TBL compared to ZPG TBL. Lehmkuhl {\it et al.} \cite{leh2020} used LES to investigate periodic control applied on the suction side of a SD7003 airfoil wing and the JAXA standard model. They found that the periodic forcing successfully eliminated the recirculation bubble in both cases. With respect to canonical flows, {\it i.e.} those simulated in a flat-plate-like configuration in absence of surface curvature, the APG that induces the TSB is commonly generated using suction-blowing (SB) or suction-only (SO) boundary conditions at the top of the domain. The SB approach has the advantage of conserving mass flux across the domain boundaries, and yields a more stable recirculation bubble caused by the favourable pressure gradient (FPG) that follows the APG. For the SO case, the flow is naturally reattached by the turbulent diffusion of momentum, so it can better resemble the separation type found in wings \cite{wu_22}. Cho {\it et al.} \cite{cho16} focused on a TSB generated by SB on the top boundary of the domain. Using periodic forcing and sweeping across a range of characteristic frequencies ($St=fL_b/U_\infty$, where $L_b$ is the uncontrolled bubble length and $U_\infty$ is the free-stream velocity), they found that low frequencies $St<0.5$ tend to reduce the separation bubble and high frequencies $St>1.56$ induce an intermittent separation/reattachment pattern. Wu {\it et al.} \cite{wu_22} investigated a TSB generated by SO and the effect of periodic control for a range of $St$. It was found that the frequency $St=0.45$ as well as a higher frequency $St=1.125$ frequency yield a 50\% reduction of the TSB. On the other hand, a much higher frequency, $St=4.5$, had no effect on the TSB.

The proliferation of computational resources has facilitated the broader application of machine learning (ML) to a wider range of problems. Notably, deep reinforcement learning (DRL), a subset of ML, proves particularly fitting to address AFC challenges and formulate effective control strategies as an alternative to the classical fluid-dynamics theory approach. In this work, we perform classical control and DRL control of an APG TBL hence comparing both control strategies. The main motivation behind the use of DRL is to allow the controlling agent certain freedom in selecting the most appropriate action for a given flow state, instead of using a single frequency as in classical periodic control. Therefore, DRL allows for an unconstrained closed loop control strategy that can adapt to the system dynamics based on previously learnt experiences. In a nutshell, DRL consists of two main entities: an agent, which is composed by a neural network (NN), and an environment, which is represented by the flow simulation in the present context. These two entities interact in a trial-an-error process in the form of episodes. This interaction occurs through three communication channels: a) The state, $s(t)$, which is sent from the environment to the agent. The state can be a total or partial observation of the environment at a given time $t$. In the case of a numerical simulation, $s$ is usually a set of probes intentionally distributed in the domain regions of interest which sample, {\it e.g.} velocity or pressure. b) The action, $a(t)$, which is sent from the agent to the environment. The action is a modification to the environment, {\it e.g.} a new boundary condition, which in computational fluid dynamics (CFD) is associated with the control strategy. For example, the action can be the mass flow rate of synthetic jet actuators, a modulation of the temperature profile in a wall, or a nudge in a turbulence model constant, among others. c) The reward, $r(t)$, which is sent from the environment to the agent. The reward is the parameter that represents the environment fitness with respect to an optimization goal, {\it e.g.} the drag coefficient if the goal is to reduce drag. Therefore, the agent will try to optimize a reward $r$ through the use of an action $a$ which is selected based on a state $s$, {\it i.e.} the agent will use a policy $\pi(a | s)$ to optimize the reward. The policy defines a probabilistic mapping from the current state $s$ to the action $a$. During training, the agent works in the so-called exploration mode by adding noise on the action sampling process. In this way, new dynamics can be explored and learnt by the algorithm. During a training episode, a collection of $\left(s,r,a\right)$ triplets is generated, also known as a trajectory. After an episode finalizes, the resulting trajectory is used to update the NN weights so that the accumulated reward expectation is maximized. Once the training is finalized, the agent is used in the so-called deterministic mode, where the most probable actions are selected (the mean of the distribution).

The applications of DRL for AFC have grown exponentially in the recent years. The reader is referred to Ref. \cite{vig23} for a wide overview of DRL for AFC, where the most representative works of different problems are discussed. Some of the most typical cases studied are drag reduction on a cylinder, both in 2D \cite{rab19_2,tan20,xu_20,li_21,var22} and 3D \cite{sua23,wan23,yan23}; convective heat reduction in Rayleigh-B\'enard problems \cite{bei20,vig23b}; reduction of the skin-friction coefficient in turbulent channels \cite{gua23, son23}; and turbulence modelling \cite{nov21,bae22,kur23}. The increasing tendency of tackling more computationally expensive cases has motivated the development of novel DRL techniques that can accelerate the training process of a control strategy. A first approach is to simulate multiple environments in parallel, also known as multi-environment DRL. With this approach, the agent trains faster by generating multiple experiences in parallel. This method has an almost perfect scaling as shown by Rabault and Kuhnle \cite{rab19}. A second approach, independent to the first one, is to use a multi-agent reinforcement learning (MARL) method.  First introduced by Belus {\it et al.} \cite{bel19} and later named by Vignon {\it et al.} \cite{vig23b}, the MARL method exploits the domain spatial invariants so that the dimensionality of the actuation space is reduced. For example, consider a set of multiple synthetic jets placed along the span of a circular cylinder. Since the flow is statistically invariant along the spanwise direction, every individual jet can be treated separately within its own span subdomain (or pseudo-environment). In this approach, rather than predicting multiple actions simultaneously, individual agents are dedicated to each pseudo-environment, each responsible for predicting a single action. The key is that all these agents share the same policy. The advantage of this approach is to reduce the number of combinations that yield an overall positive action, hence avoiding the curse of dimensionality. As reported in Refs. \cite{sua23} and \cite{vig23b}, using MARL finally allowed the agent to learn the system dynamics and provide positive actions. Defining the number of parallel environments as $N_\mathrm{e}$ and the number of parallel pseudo-environments within each environment $N_\mathrm{pe}$, a total of $N_\mathrm{e}N_\mathrm{pe}$ trajectories can be sampled in parallel, and the training time of an agent can be greatly reduced.

With the aim of investigating classical periodic control and DRL control on a TSB, the paper has been structured as follows: \S\ref{sec:methodology} introduces the mathematical framework for the simulation and the different control strategies. \S\ref{sec:results} presents and discusses results for classical control in \S\ref{sec:classic_results}, and DRL control in \S\ref{sec:drl_results}. Finally, conclusions are presented in \S\ref{sec:conclusions}.

\section{Methodology}
\label{sec:methodology}
\vspace*{0.25cm}
\subsection{CFD setup}
\vspace*{0.25cm}
Following the LES method, the spatially filtered incompressible non-dimensional Navier--Stokes (NS) equations
\begin{align}
    \nabla\cdot\overline{\boldsymbol u}&=0\\
    \partial_t\overline{\boldsymbol u}+\left(\overline{\boldsymbol u}\cdot\nabla\right)\overline{\boldsymbol u}&=-\nabla\overline{p}+Re^{-1}\nabla^2\overline{\boldsymbol u} - \nabla\cdot\boldsymbol\tau,
\end{align}
are numerically solved on a discrete domain, where $\overline{\boldsymbol u}=(\overline{u},\overline{v},\overline{w})$ is the filtered velocity vector field, $\overline{p}=\overline{P}/\rho$ is the density-scaled filtered pressure, and $\boldsymbol\tau=\overline{\boldsymbol u\otimes\boldsymbol u}-\overline{\boldsymbol u}\otimes\overline{\boldsymbol u}$ is the sub-grid scale (SGS) stress tensor.
The deviatoric part of the SGS tensor is modelled using the Boussinesq hypothesis
\begin{equation}
    \boldsymbol{\tau}^d = \boldsymbol{\tau} - \frac{2}{3}k\boldsymbol\delta = -2 \nu_{\mathrm{sgs}} \overline{\boldsymbol{S}},
\end{equation}
where $k=\mathrm{tr}(\boldsymbol\tau)/2$ is the turbulent kinetic energy, $\boldsymbol{\delta}$ is the Kronecker delta, and $\overline{\boldsymbol{S}}=\left(\nabla\otimes\overline{\boldsymbol{u}}+\overline{\boldsymbol{u}}\otimes\nabla\right)/2$ is the rate-of-strain tensor. The SGS viscosity is finally closed with the Vreman model \cite{vre04}. The overbar notation of the LES filtering operation for flow-field quantities is thereafter implied.

The high-order spectral-element-method solver \href{https://gitlab.com/bsc_sod2d/sod2d_gitlab}{SOD2D} \cite{gas23} is used to simulate an incompressible APG TBL in a computational domain with dimensions $L_x\times L_y\times L_z=1100\times120\times125$, corresponding to the streamwise $x$, wall-normal $y$, and spanwise $z$ directions. A coarse grid and a well-resolved (fine) grid composed of 5\textsuperscript{th}-order hexahedral elements are considered, and details are given in table \ref{tab:grids}. The idea behind considering 2 different resolutions is to train the DRL model on the coarse grid, thus significantly reducing the computational cost of training. Afterwards, the model can be trained further on the fine grid via transfer learning, {\it i.e.} using the optimized NN weights already available as the training starting point on the fine grid. Alternatively, given that similar dynamics are captured in both coarse and fine grids, the controlling agent can be directly applied into the fine-grid simulation without further training.

\begin{figure}[h]
    \centering
    \includegraphics[width=0.9\textwidth]{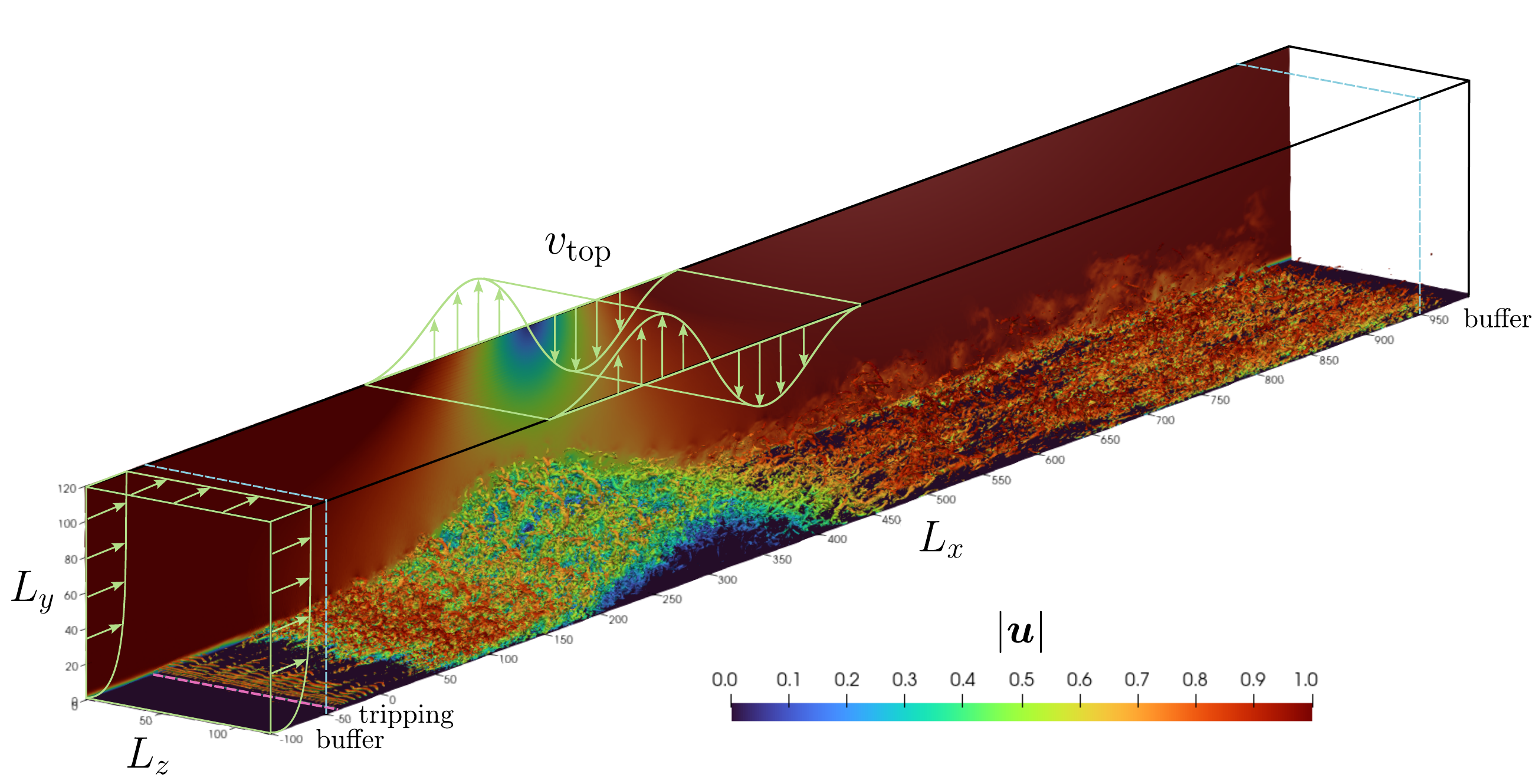}
    \caption{Schematic representation of the computational domain together with an instantaneous snapshot of vortical structures captured by the Q criterion and colored by velocity magnitude.}
    \label{fig:computational_domain}
\end{figure}

With respect to the numerical methods implemented in the flow solver, an implicit-explicit Adams--Bransford Crank--Nicolson scheme is used together with the fractional-step method to solve the governing equations. In addition, a matrix free conjugate gradient preconditioned by the diagonal is used to solve the linear part of the NS equations and the pressure equation. The computational domain sets a buffer zone of 50 units in the $x$ direction at the inlet and at the outlet, yielding an effective streamwise domain of 1000 units ($x\in[-50,950]$). Dimensions are normalized by the displacement thickness at the inlet $\delta^*_0$, and the velocity field is normalized by the streamwise velocity imposed at the top of the domain $U_\infty$. The flow is initialized using a Blasius boundary layer at $Re_{\delta_0^*} = 450$, where $Re_{\delta_0^*} = U_\infty \delta_0^*/\nu$ is the Reynolds number based on the displacement thickness. The inlet boundary condition retains the Blasius profile used as initial condition. At $x=-40$, the laminar boundary layer is tripped using the method explained in Refs. \cite{sch12, hos16} to accelerate the transition to turbulence. On top the domain, the APG is defined using SB similar to Refs. \cite{abe17, wu_20}
\begin{equation}
    v_\mathrm{top} = v_\mathrm{max}\sqrt{2}\left(\frac{x_c-x}{\sigma}\right)\exp\left[\psi-\left(\frac{x_c-x}{\sigma}\right)^2\right],
\end{equation}
where $v_\mathrm{max}=0.4$, $x_c=306.64$, $\sigma=110.49$, and $\psi=0.95$. We note that $v_\mathrm{max}$ is 20\% larger than the value selected by Wu {\it et al.} \cite{wu_20} so that the TSB is still formed when using the coarse LES grid. Moreover, at the top of the domain, a zero spanwise vorticity condition ($\omega_z = 0$) and a homogeneous Neumann condition for the spanwise velocity ($\partial_y w=0$) are applied. A periodic boundary condition is imposed in the spanwise direction, and a convective outlet is set on the streamwise end of the domain. On the wall, the lower part of the domain, a classical non-slip boundary condition is imposed. Finally, in the outlet, a pressure-based boundary condition is applied together with the sponge zone. In the fine grid, the a-posteriori computation of the wall-shear stress yields an approximate friction Reynolds number of $Re_\tau=180$ and $Re_\tau=750$ at $x=150$ and $x=900$, hence right upstream of the TSB and right before the convective outlet of the domain, respectively. A schematic representation of the computational domain which summarizes the setup is shown in figure \ref{fig:computational_domain}, where vortical structures represented with the Q criterion \cite{hun88} are depicted too.

\begin{table}[h]
    \begin{center}
        \begin{tabular}{lccccc}
        \br
         &$\Delta x^+$&$\Delta z^+$&$\Delta y^+|_\mathrm{min}$&$\Delta y^+|_\mathrm{max}$&DOFs\\
        \mr
        Coarse &38.20&38.20&1.27&25.08&4,063,913\\
        Fine &15.28&15.28&0.93&14.51&40,543,349\\
        \br
        \end{tabular}
    \end{center}
    \caption{Grid details. $\Delta$ refers to grid spacing, the $\cdot^+$ notation refers to viscous units (scaling with $u_\tau$ and $\nu$), and DOFs stands for degrees of freedom, {\it i.e.} the total number of nodes in the spectral-element domain discretization.}
    \label{tab:grids}
\end{table}

Once the flow is initialized, the simulation is run until the TSB is formed and the flow has fully developed. Afterwards, the control surfaces start acting on the flow. There are $N_\mathrm{ac}=6$ rectangular actuators located at $x\in[150,195]$, right upstream of the separation bubble, and each actuator has a spanwise width of $d_\mathrm{ac}=10.42$ which is also the length between adjacent actuators. A classical control strategy similar to those reported in Refs. \cite{wu_22,cho16} is considered. Classical control is denoted as the harmonic time-forcing of the wall-normal velocity, defined as
\begin{equation}
    v_\mathrm{ac}=A_\mathrm{ac}\sin\left(2\pi f_\mathrm{ac}t\right).
\end{equation}
By definition, the free parameters in classical control are the actuation amplitude $A_\mathrm{ac}$ and the actuation frequency $f_\mathrm{ac}$. Depending on the APG top-boundary condition type, the actuation amplitude needs to be adjusted. We use $A_\mathrm{ac}=0.3U_\infty$, which corresponds to a momentum-flux coefficient of $C_\mu=(v_\mathrm{ac, rms}/U_\infty)^2 N_\mathrm{ac}d_\mathrm{ac}/L_z=2.25\%$, where `rms' stands for root mean square, and it is within the effective range $0.01\%<C_\mu<3\%$ \cite{cho16,gre00}. We note that Wu {\it et al.} \cite{wu_22} used $C_\mu=0.0625\%$ in a SO APG setup, and Cho {\it et al.} \cite{cho16} used $C_\mu=1.4\%$ in a SB APG setup, even though the top wall-normal velocity profile spanned the entire streamwise length of the domain in the latter case. The SB APG setup requires a larger actuation amplitude because of the forced reattachement arising from the FPG that follows the APG. On the other hand, sizing of the control surfaces is based on the objective of effectively modulating the low-frequency motion inherent in the uncontrolled flow. These motions are closely associated with the Görtler instability, which exhibits a spanwise wavelength of the same order of magnitude than the boundary-layer thickness and extends by more than $10$ times the boundary-layer thickness in the streamwise direction \cite{wu_20}. More details and comparisons with other orientations of the control surfaces or rounded actuators can be read in Refs. \cite{ara11,smi02,bur16} and \cite{les11,dan06,wu_09}, respectively.

\begin{figure}[!t]
    \centering
    \begin{subfigure}[t]{.45\linewidth}
        \centering
        \includegraphics[width=\linewidth]{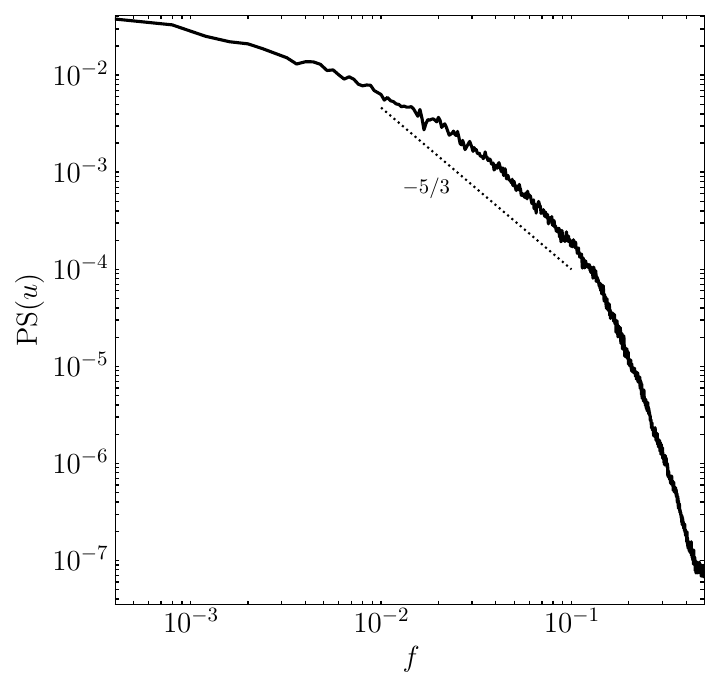}
    \end{subfigure}
    \begin{subfigure}[t]{.45\linewidth}
        \centering
        \includegraphics[width=\linewidth]{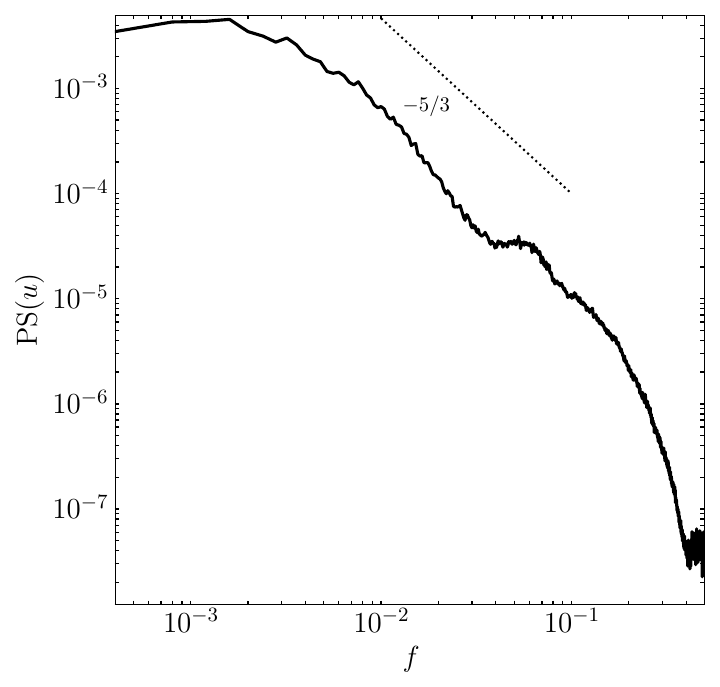}
    \end{subfigure}
    \caption{Power spectrum (PS) of the streamwise velocity component $u$ sampled at 2 different locations. Left: $(x,y)=(324, 15)$, corresponds to a point inside the TSB. Right: $(x,y)=(372, 55)$, corresponds to a point near the downstream shear layer of the TSB. The PS is computed using the Welch method for a temporal signal of 15000 time units (TU) in total, split into 6 segments with 50\% overlap. Additionally, 6 PS are computed along $z$ for each $(x, y)$ which are then averaged and result in the displayed spectra.}
    \label{fig:spectra_nonactuated_fine}
\end{figure}

In the current setup, an actuation frequency of $f_\mathrm{ac}=0.0025$ is selected (normalized by $\delta^*_0$ and $U_\infty$). This is the same frequency as identified by Wu {\it et al.} \cite{wu_20} for both SO-TSB and SB-TSB configurations (named the high frequency). Wu {\it et al.} \cite{wu_22} also selects this frequency for periodic control in a SO configuration. In this direction, we perform a spectral analysis of the streamwise velocity component at different locations. In figure \ref{fig:spectra_nonactuated_fine}, the spectra of a probe inside the TSB and another probe near the TSB downstream shear layer are displayed. While the downstream probe reaches a peak near $f=0.0015$, a clear dominant frequency cannot be detected. This process is repeated for all the probes displayed in figure \ref{fig:wit_top} and the $v, w, p$ flow fields (not shown here for the sake of brevity), yielding a similar conclusion. On the other hand, Cho {\it et al.} \cite{cho16} observes that the most effective frequency for the reduction of the TSB in a SB setup is $St=0.5$. In this work, the non-actuated separation bubble has an approximate length of $L_b=143$ (more results on the next section), which translates our selected actuation frequency to $St=0.3575$, still within the range of effective actuation frequencies proposed by Cho {\it et al.} \cite{cho16}.

\vspace*{0.25cm}
\subsection{DRL setup}
\begin{figure}[h]
    \centering
    \includegraphics[width=0.9\textwidth]{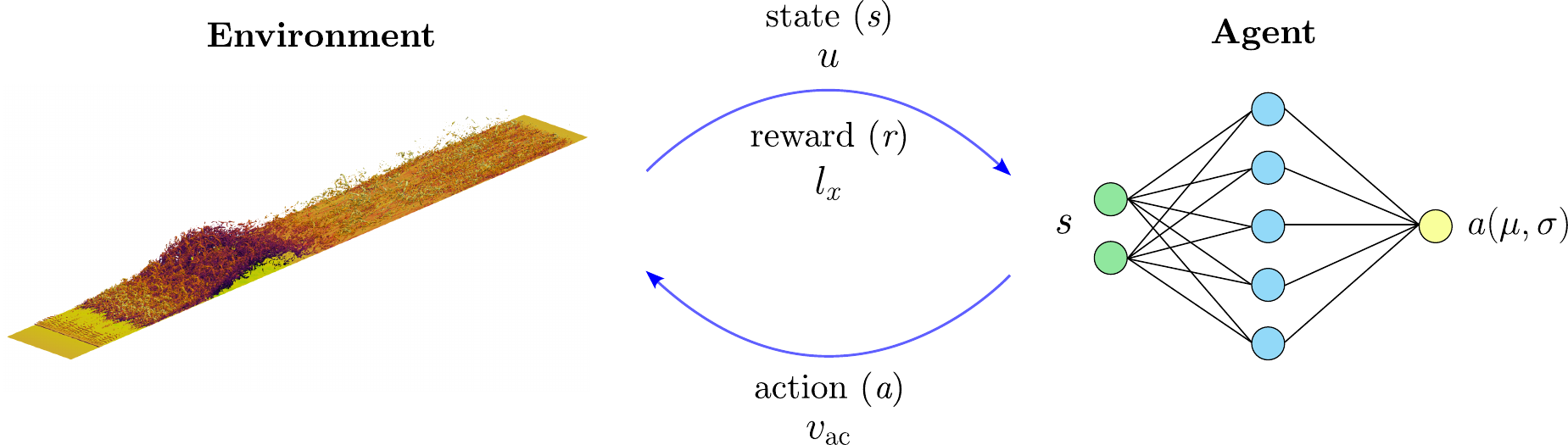}
    \caption{CFD--DRL setup. Trajectories consisting of state-reward-action triplets $\left\{\left(s^0,r^0,a^0\right),\left(s^1,r^1,a^1\right),...,\left(s^n,r^n,a^n\right)\right\}$ are sampled during an episode. The NN weights are then optimized based on the trajectories to find the best action distribution that maximizes the accumulated reward in time.}
    \label{fig:rl_scheme}
\end{figure}

\begin{figure}[h]
    \centering
    \includegraphics[width=0.7\textwidth]{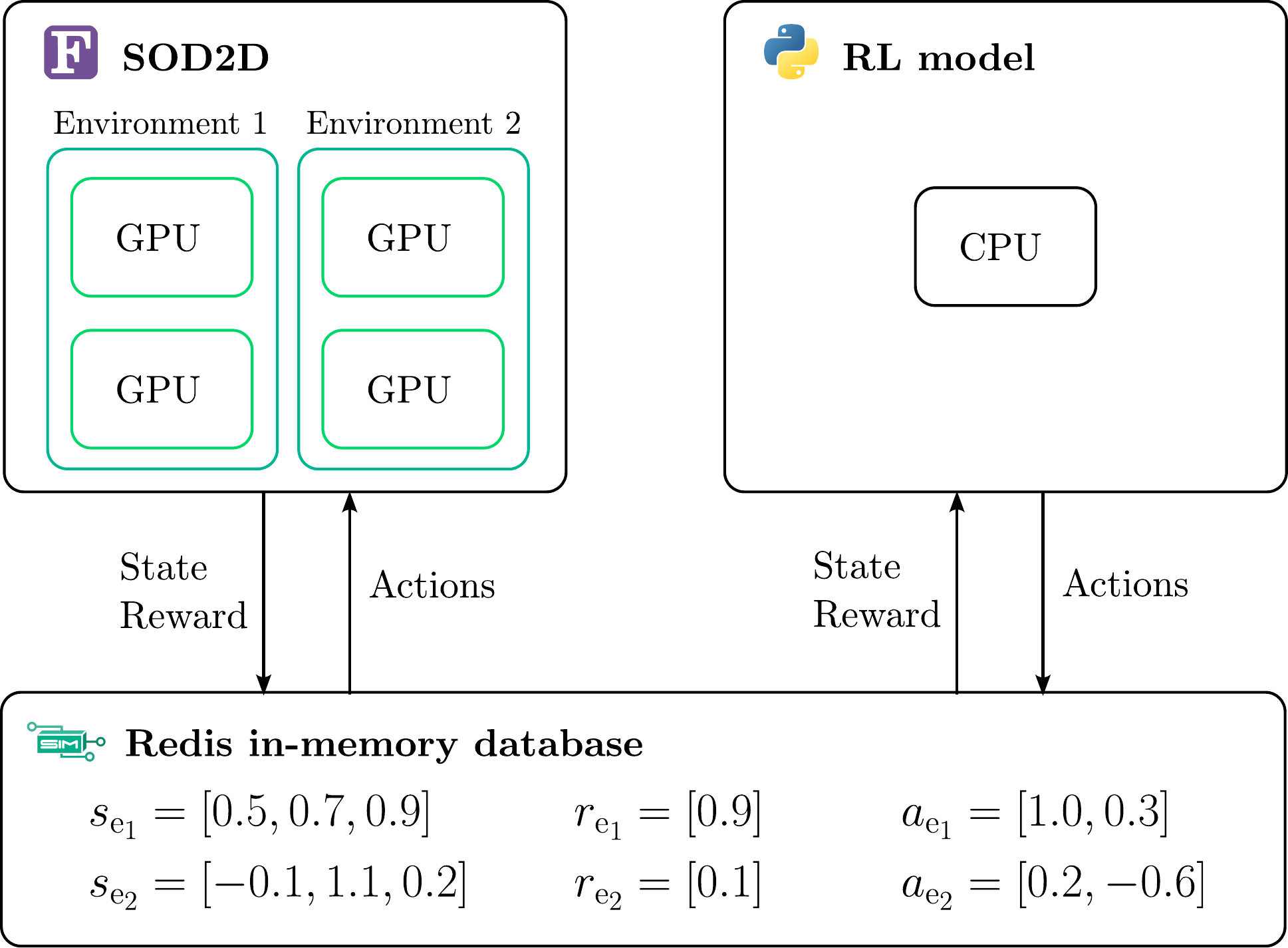}
    \caption{Communication scheme between the multiple CFD environments run in parallel using GPUs and the DRL model run on a CPU. The in-memory Redis database handles the communication of state, reward, and actions across the solver and the model. For the sake of simplicity, only 2 parallel CFD environments are represented, but we typically run up to 8 parallel CFD environments on the cluster.}
    \label{fig:comms}
\end{figure}

\vspace*{0.25cm}
Beyond periodic forcing, we also consider DRL control to reduce the TSB length. As we have briefly summarized in the introduction, in a DRL problem we define two main entities: the environment, {\it i.e.} the system where we will apply the control actions, and the agent, {\it i.e.} the model in charge of predicting actions. In the current framework, the environment is the simulation performed by the CFD solver, and the agent is a NN that predicts a probability distribution of possible actions. This is depicted schematically in figure \ref{fig:rl_scheme}. The software used to simulate the environment is again the SOD2D CFD solver, and the \href{https://github.com/tensorflow/agents}{TF-agents} \cite{tfa18} is used for the DRL model part. A challenge that commonly arises when linking high-performance physics solvers (typically written in Fortran/C/C++) and high-level libraries that implement ML models (typically Python) is the communication between the different executable instances, also known as the two-language problem. While this can be accomplished through Unix sockets \cite{fon21} or message-passing interface (MPI) \cite{gua23}, in the current setup the \href{https://github.com/CrayLabs/SmartSim}{SmartSim} \cite{par22} library is employed. SmartSim allows communication between processes through an in-memory Redis database with minimal overhead and, compared to the previously mentioned approaches, it lowers the software complexity of the framework. Other projects such as \href{https://github.com/flexi-framework/relexi}{RELEXI} \cite{kur23,kur22} have already successfully used SmartSim to handle communication across CFD solver and DRL model. In the current framework, the main computation workload arises from the CFD temporal integration, so the SOD2D solver is run on graphics-processing units (GPUs) in parallel. On the other hand, the DRL model is based on a small multi-layer perceptron NN and this can be easily handled by the central-processing unit (CPU) part of a cluster node which would remain idle otherwise. A schematic diagram of this setup is depicted in figure \ref{fig:comms}, noting the multi-environment approach as introduced earlier. The described framework is implemented in the \href{https://github.com/b-fg/SmartSOD2D}{SmartSOD2D} package which links SOD2D and TF-agents through SmartSim in a high-level abstraction. We also note that all this software is open-source and freely available to the reader.

The same actuators as defined for the periodic control are employed in the DRL framework. However, to impose mass conservation both in space and time, we group the actuators in pairs and the DRL model sets a mass flux value on one actuator, and the opposite value on the other actuator. As explained before, the MARL approach can be used when the flow is invariant in a certain direction. In this case, the flow is invariant in the spanwise direction, and a subdomain composed by one pair of actuators is defined as a pseudo-environment. Therefore, a total of $N_\mathrm{pe}=3$ pseudo-environments are defined for each CFD environment hence generating 3 different trajectories that the agent uses during the optimization step. Also, the output dimensionality is reduced from $3$ to $1$, hence reducing the number of combinations needed to explore all possible control strategies. With the aim of equating the classical and DRL controls, we allow the agent to explore actions with a maximum absolute value equal to the amplitude set in the classic control, $|v_\mathrm{ac}|_\mathrm{max}=0.3$. Furthermore, an exponential smoothing function is applied to the discrete actions predicted by the agent, hence forcing a smooth continuous control signal applied at every time step in the environment.

\begin{figure}[h]
    \centering
    \begin{subfigure}[t]{1\linewidth}
        \centering
        \includegraphics[width=\linewidth]{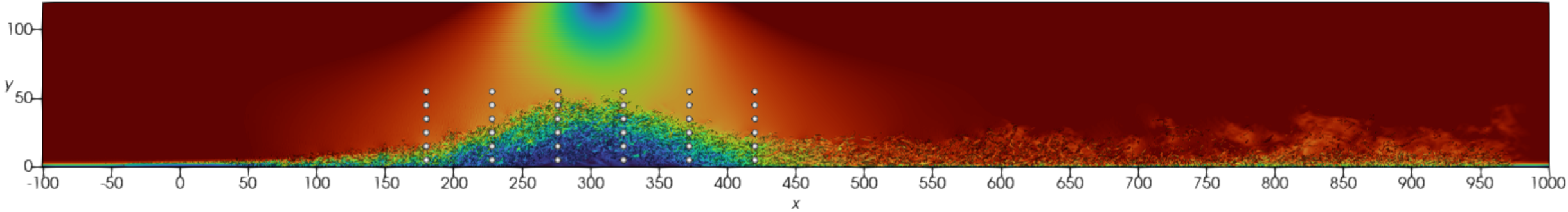}
    \end{subfigure}
    \begin{subfigure}[t]{1\linewidth}
        \centering
        \includegraphics[width=1\linewidth]{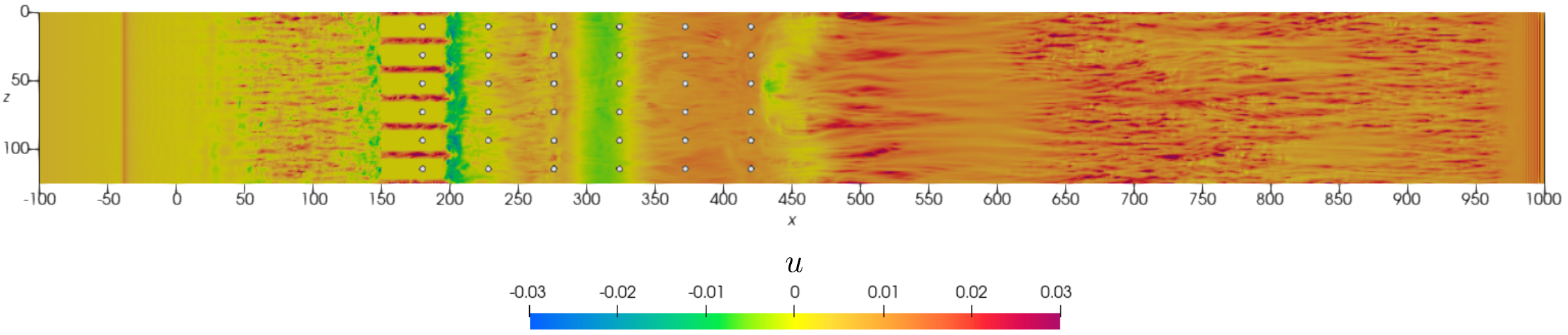}
    \end{subfigure}
    \caption{Distribution of witness points. Top: non-actuated $xy$-plane at $z=L_z/2$ displaying the same flow-field information (and colormap) as in figure \ref{fig:computational_domain}. Bottom: periodic-forcing $xz$-plane of the first off-wall node displaying the streamwise $u$ velocity component.}
    \label{fig:wit_top}
\end{figure}

The environment state $s$ is defined as a cloud of witness points (probes) that measure the streamwise velocity $u$. The witness points cloud is composed by an equidistant $6\times6\times6$ points grid spanning a $240\times50\times104$ subdomain that captures the non-actuated recirculation bubble region and its vicinity, as depicted in figure \ref{fig:wit_top}. Since the MARL approach is used and the computational domain is divided into 3 pseudo-environments, it is important to use a representative state for each subdomain. Therefore, the state of each pseudo-environment is composed of $2$ planes of witness points aligned with the actuators in the streamwise direction totaling 72 witness points.

The environment reward $r$ is based on the recirculation length of the turbulent bubble, and the optimization process aims to maximize $r$ so that the recirculation area is reduced. To calculate a characteristic length for the recirculation bubble ($l_x$), the wall-shear stress value is computed at each element face of the wall surface. The area of those elements with a negative wall-shear stress $\left(A_i|_{\tau_w<0}\right)$, {\it i.e.} with local recirculation, is integrated and then divided by the span of the pseudo-environment
\begin{equation}
    l_x = \frac{N_\mathrm{pe}}{L_z}\sum_{i}A_i|_{\tau_w<0}.
\end{equation}
The reward is normalized by the time-averaged non-actuated characteristic recirculation length $\overline{l_x^*}$, yielding $r=-l_x/\overline{l_x^*}$ (noting that the overline notation is now related to temporal averages).
Additionally, instead of using an instantaneous value, the reward is averaged during an actuation period so that the overall response of the system to a given action is more representative. Last, each pseudo-environment computes a local reward $r_l$ and the global reward $r_g$ is an average of the local rewards. These two quantities are finally merged in a weighted sum, $r=\alpha r_l + (1-\alpha)r_g$, where $\alpha$ is a tunable parameter, and the $\alpha=0.5$ value is selected because it represents an unbiased trade-off between both rewards.

With respect to the DRL model, the proximal policy optimization (PPO) \cite{sch17} algorithm is employed as the controlling agent. The PPO method is a policy-gradient method which tries to find an optimal policy, {\it i.e.} distribution of actions, which maximizes the accumulated reward expectation given the state of the environment. The main advantages of PPO with respect to other DRL algorithms are its small number of tunable parameters and its suitability for continuous control problems \cite{rab19_2,mni15}. In the current framework, we employ a NN consisting of 2 layers with 128 neurons per layer. An episode duration of $T_e=4/f_\mathrm{ac}=1600$ is set (4 periods of the periodic-forcing frequency) so that the low frequencies of the system can develop. The actuation frequency of the DRL model is $f_\mathrm{ac,DRL}=10f_\mathrm{ac}$ yielding 40 actuations per episode, which we considered a trade-off between actuating too often (the flow has not enough time to develop after a new actuation) and actuating too seldom (the flow is not acted on when required). Actuating every 10\% of the system's lowest frequency is within the order of magnitude that can be found in the literature \cite{rab19_2}. We also note that 8 environments are typically run in parallel depending on the cluster availability and architecture. Given that 3 pseudo-environments are defined within each CFD simulation, this results in 24 trajectories sampled in parallel.

\section{Results and discussion}
\label{sec:results}
\vspace*{0.25cm}
\subsection{Periodic control}
\label{sec:classic_results}
\vspace*{0.25cm}

\begin{figure}[h]
    \centering
    \begin{subfigure}[t]{1\linewidth}
        \centering
        \includegraphics[width=\linewidth]{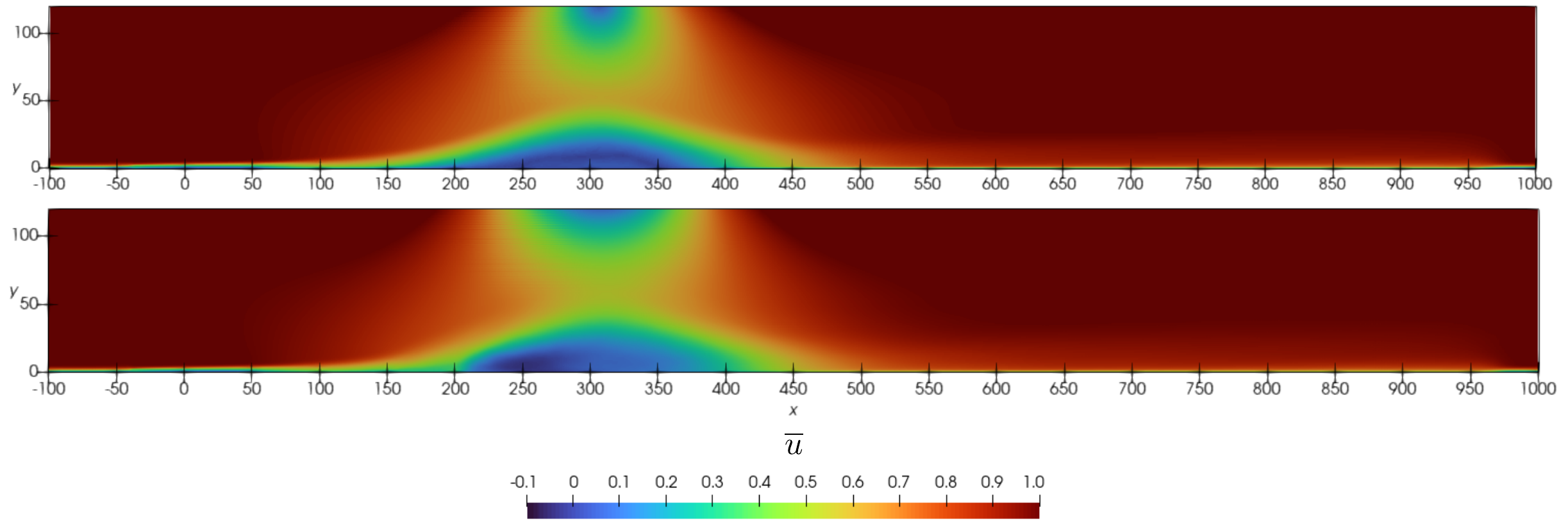}
        \caption{Time-averaged streamwise velocity component $\overline{u}$ at $z=L_z/2$.}
    \end{subfigure}
    \begin{subfigure}[t]{1\linewidth}
        \vspace*{0.2cm}
        \centering
        \includegraphics[width=1\linewidth]{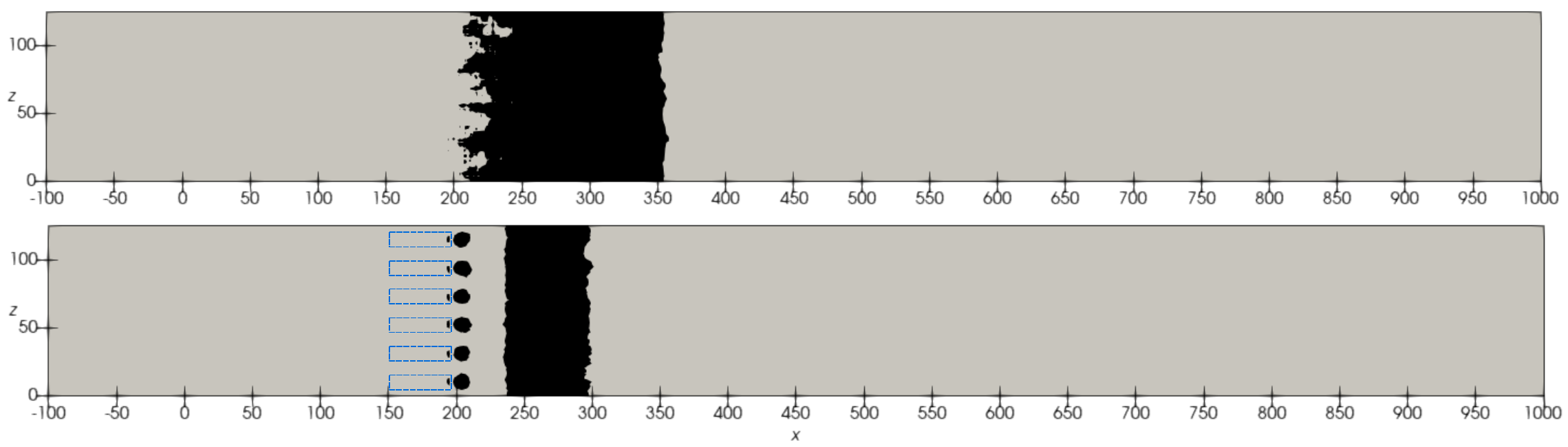}
        \caption{Time-averaged recirculation region ($\overline{u}<0$) at the $xz$-plane of the first off-wall node.}
    \end{subfigure}
    \caption{Fine-grid time-averaged flow field. Both panels (a) and (b) show non-actuated (top) and periodic control (bottom) results.}
    \label{fig:front_fine}
\end{figure}

\begin{figure}[!t]
    \centering
    \begin{subfigure}[t]{.45\linewidth}
        \centering
        \includegraphics[width=\linewidth]{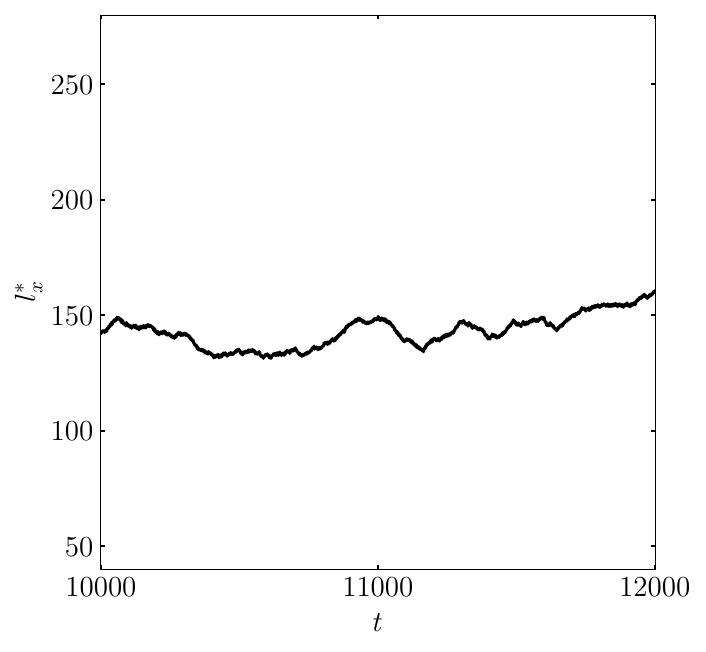}
    \end{subfigure}
    \begin{subfigure}[t]{.45\linewidth}
        \centering
        \includegraphics[width=1\linewidth]{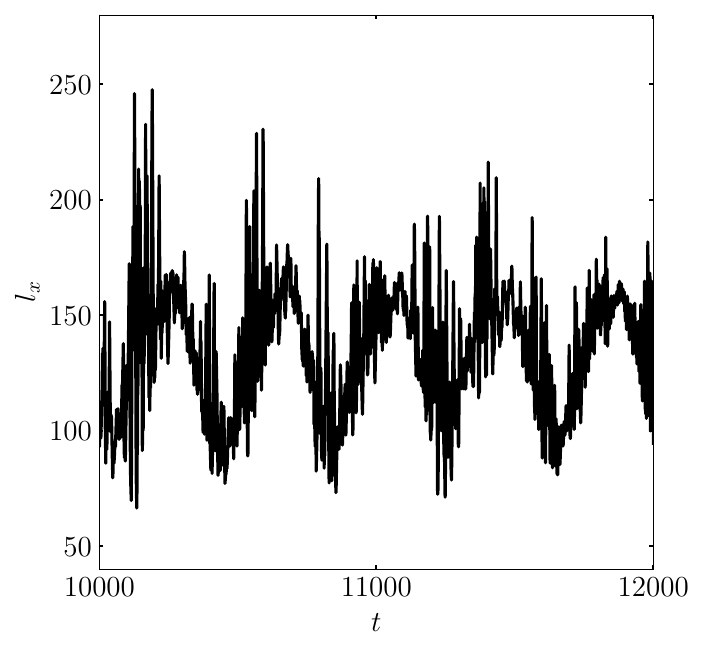}
    \end{subfigure}
    \caption{Temporal signal of the characteristic recirculation length for the non-actuated (left) and periodic control (right) fine-grid cases.}
    \label{fig:lx}
\end{figure}

Results of the classical time-periodic forcing control are first compared with the non-actuated TSB. The open-loop control starts actuating after 5000 time units (TU, normalized by $\delta^*_0$ and $U_\infty$) and statistics are recorded during 15000TU for both the coarse and fine grids. Figure \ref{fig:front_fine} displays the time-averaged streamwise velocity $\overline{u}$ for the fine-grid case. It is observed that the non-actuated case exhibits a recirculation bubble generated by the SB top-boundary condition that qualitatively spans from $x=225$ to $x=350$. On the other hand, the actuated bubble qualitatively spans a smaller region of the domain and new separation bubbles are spotted near the actuators. The reduction of the TSB on its downstream region is a phenomenon also observed by Cho {\it et al.} \cite{cho16} in a similar SB APG TBL setup. The bubble length can be better quantified by the time-averaged characteristic recirculation length $\overline{l_x}$. As presented in table \ref{tab:lx}, a reduction of 6.8\% in $\overline{l_x}$ is obtained for the fine grid when using periodic control. It can be noted that the normalized standard deviation of $\overline{l_x}$ is similar to the rms value of the forcing signal, {\it i.e.} $\sigma(\overline{l_x})/\overline{l_x}\sim A_{\mathrm{ac}}/\sqrt{2}$, and this means that the periodic-forcing control dominates the TSB. This phenomenon can be visualized in figure \ref{fig:lx}, where the temporal signals of the non-actuated and actuated $l_x$ are shown. Because of the FPG present in the SB setup, the reattachment point of the TSB has less freedom compared to a SO setup. In the latter case, a reduction of the TSB up to 50\% can be observed when using periodic control under the correct actuation frequency, as reported by Wu {\it et al.} \cite{wu_22}.

\begin{table}[h]
    \begin{center}
        \begin{tabular}{lcccc}
        \br
        & \thead{$\overline{l_x^*}$} & \thead{$\overline{l_x}$} & $1-\overline{l_x}/\overline{l_x^*}$\\
        \mr
        Fine & $143.0\pm8.6$ & $133.3\pm27.2$ & $6.8\%$ \\
        Coarse & $153.6\pm8.3$ & $129.5\pm33.4$ & $15.7\%$ \\
        \br
        \end{tabular}
    \end{center}
    \caption{Time-averaged characteristic length $\overline{l_x}$ and standard deviation of the non-actuated and the periodic-forcing cases for both fine and coarse grids. The reduction of the characteristic recirculation length is also shown.}
    \label{tab:lx}
\end{table}

\begin{figure}[h]
    \centering
    \includegraphics[width=\textwidth]{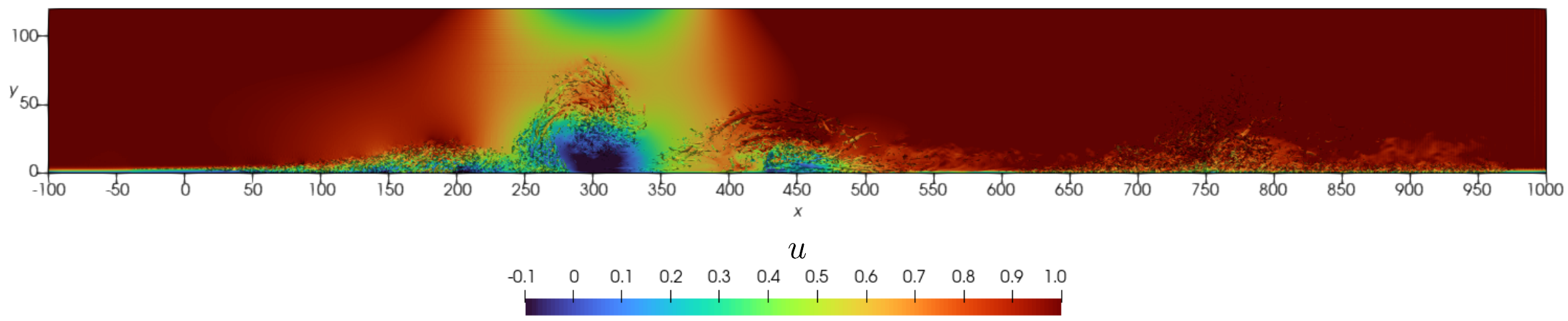}
    \caption{Instantaneous Q criterion isosurface (defined in figure \ref{fig:computational_domain}) colored by the streamwise velocity component $u$ on the periodic-forcing fine-grid case.}
    \label{fig:front_act_insta_fine}
\end{figure}

\begin{figure}[!t]
    \centering
    \begin{subfigure}[t]{.45\linewidth}
        \centering
        \includegraphics[width=\linewidth]{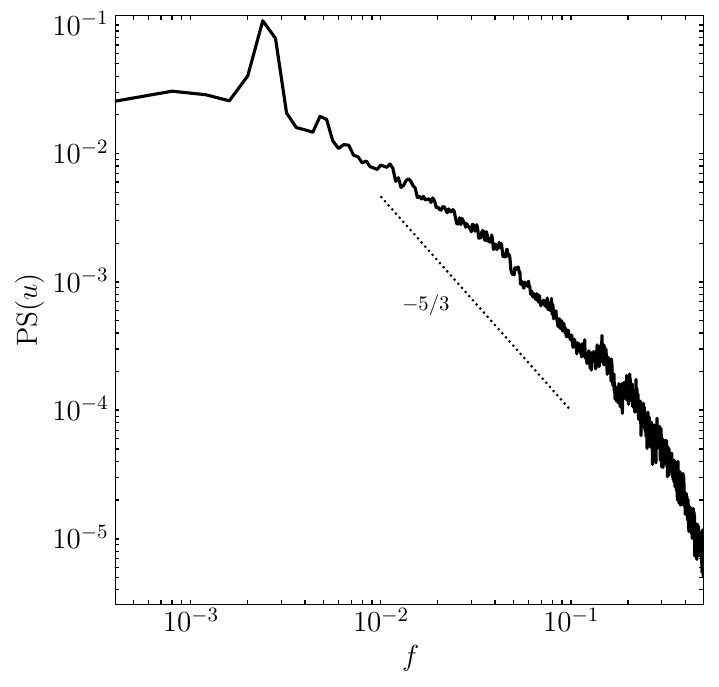}
    \end{subfigure}
    \begin{subfigure}[t]{.45\linewidth}
        \centering
        \includegraphics[width=\linewidth]{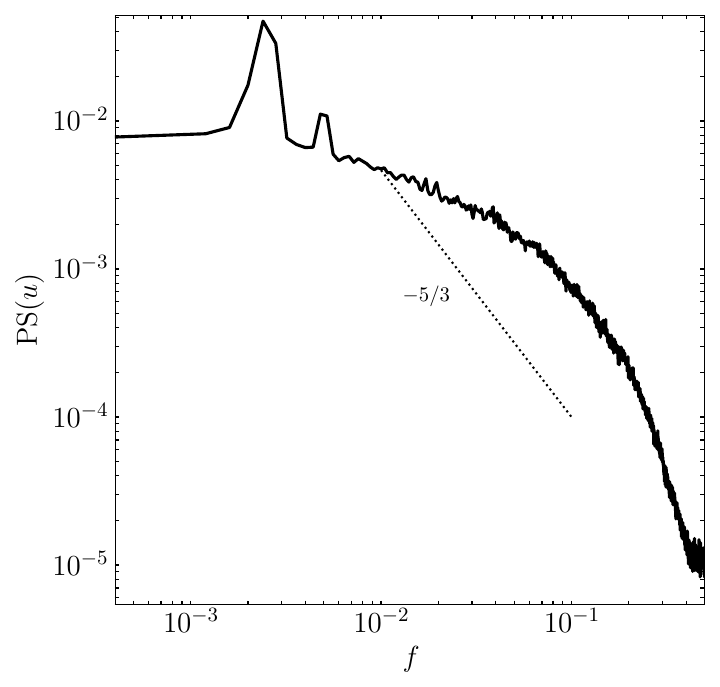}
    \end{subfigure}
    \begin{subfigure}[t]{.45\linewidth}
        \centering
        \includegraphics[width=\linewidth]{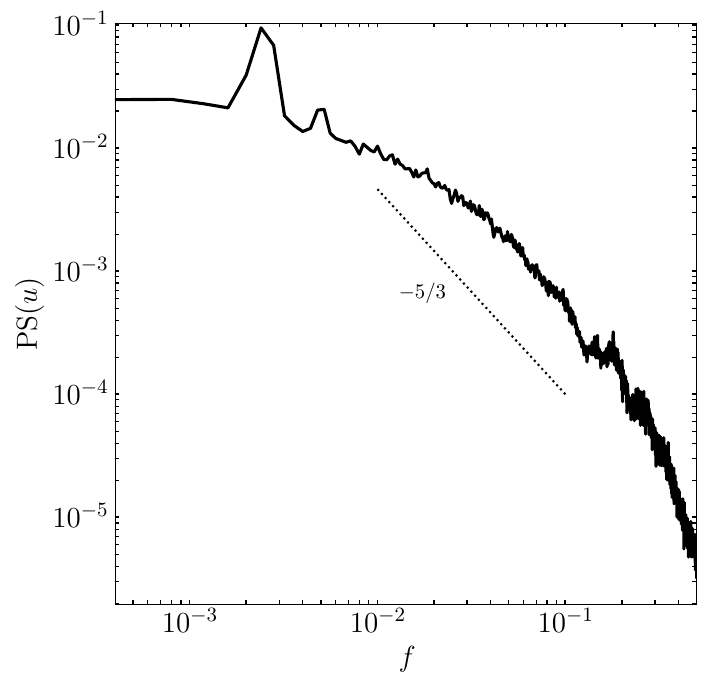}
    \end{subfigure}
    \begin{subfigure}[t]{.45\linewidth}
        \centering
        \includegraphics[width=\linewidth]{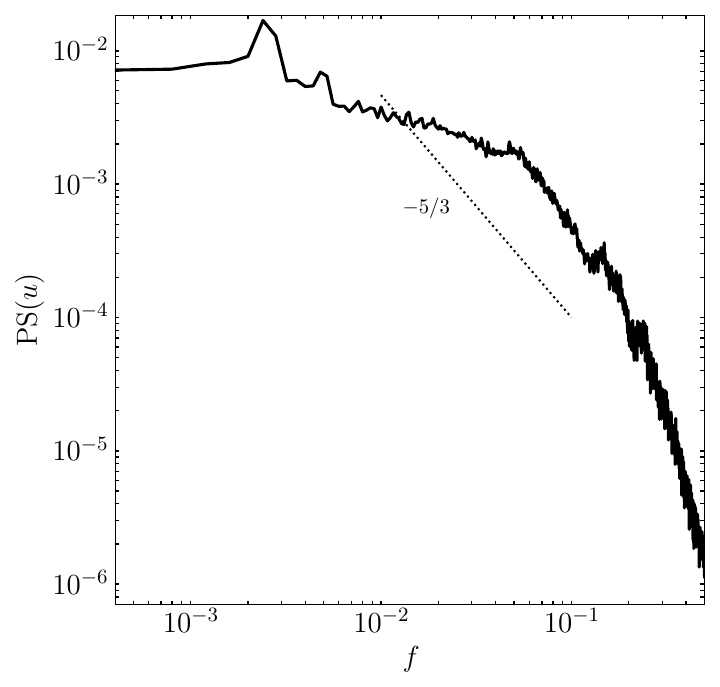}
    \end{subfigure}
    \caption{Power spectrum (PS) of the streamwise velocity component $u$ of the periodic-forcing TSB sampled at 2 different locations. Left: $(x,y)=(324, 15)$. Right: $(x,y)=(372, 55)$. Top: fine-grid results. Bottom: coarse-grid results. See figure \ref{fig:spectra_nonactuated_fine} caption for additional information on then calculation of the PS.}
    \label{fig:spectra_actuated}
\end{figure}

The instantaneous flow field of the actuated TSB is depicted in figure \ref{fig:front_act_insta_fine}. A large-scale spanwise vortical structure can be observed as a result of the harmonic actuation. The interaction of these structures with the TSB effectively reduces the size of the bubble. Additionally, the spectra of the witness points previously shown for the non-actuated case (figure \ref{fig:spectra_nonactuated_fine}) is now displayed for the periodic forcing case in figure \ref{fig:spectra_actuated}. The effect of the actuators is clearly appreciated on the peak arising precisely at $f_\mathrm{ac}$, while an harmonic peak is also present at a higher frequency.

\begin{figure}[h]
    \centering
    \begin{subfigure}[t]{\linewidth}
        \centering
        \includegraphics[width=\linewidth]{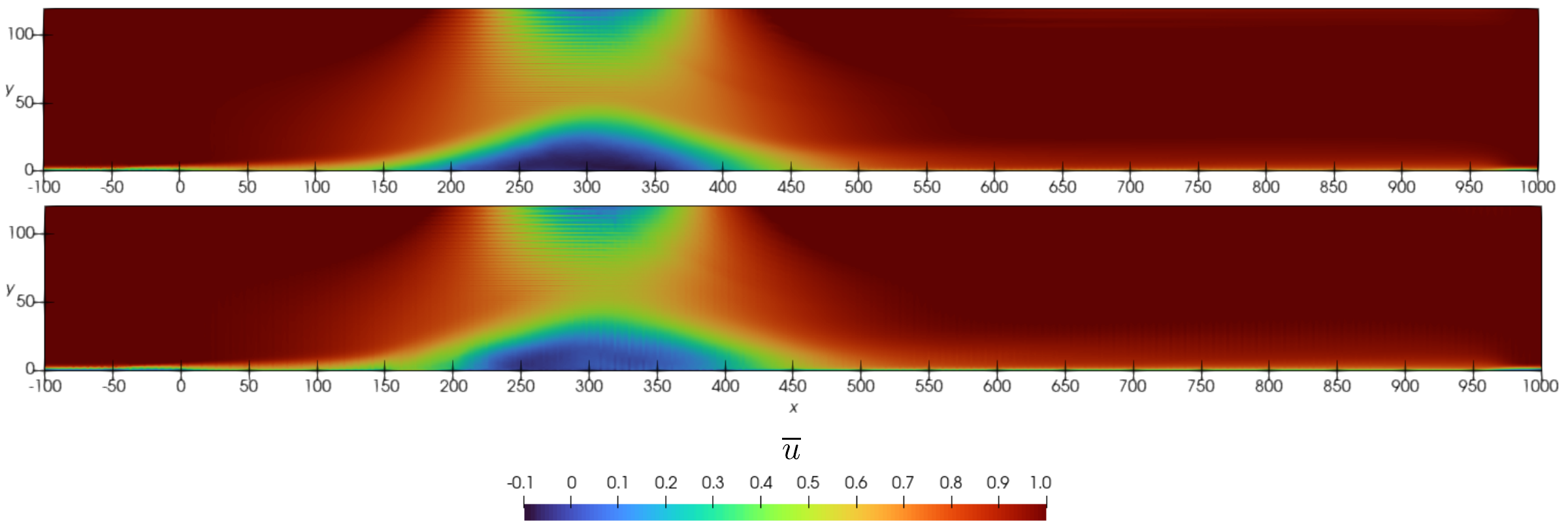}
        \caption{Time-averaged streamwise velocity component $\overline{u}$ at $z=L_z/2$.}
    \end{subfigure}
    \begin{subfigure}[t]{\linewidth}
        \vspace*{0.2cm}
        \centering
        \includegraphics[width=\linewidth]{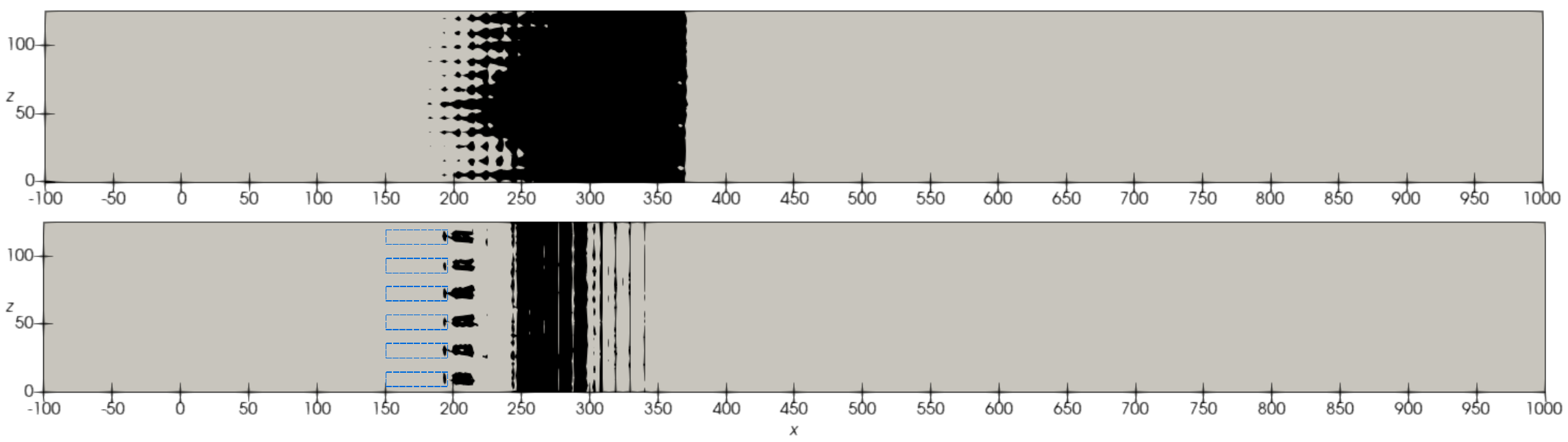}
        \caption{Time-averaged recirculation region ($\overline{u}<0$) at the $xz$-plane of the first off-wall node.}
    \end{subfigure}
    \caption{Coarse-grid time-averaged flow field. Both panels (a) and (b) show non-actuated (top) and periodic control (bottom) results.}
    \label{fig:front_coarse}
\end{figure}

The coarse-grid results are discussed next. The main objective of the coarse--fine grid comparison is to assess whether the main flow features are preserved so that the coarse grid can be used for training the DRL model. Coarse-grid results depicting the time-averaged flow field and its recirculation region are displayed in figure \ref{fig:front_coarse}. In general, a very similar flow structure compared with the fine-grid results can be observed. It can also be noted that the TSB is slightly larger in this case, and this is quantified in table \ref{tab:lx}. The characteristic recirculation length is reduced by 15.7\% when applying periodic actuation on the coarse grid. Similarly to the fine grid, the standard deviation of this metric is significantly increased as a result of the periodic forcing. Spectra of the coarse-grid simulation are also shown in figure \ref{fig:spectra_actuated}. While the general trend is in good agreement with the fine-grid results, the spectrum of the probe located downstream of the bubble and farther away from the wall (bottom right) presents some discrepancies. This can be expected since the stretching of the coarse grid significantly reduces the resolution far from the wall. Nevertheless, the main flow features of the fine-grid simulation are preserved, and we conclude that the coarse grid is adequate for training the DRL model.

\subsection{DRL control}
\label{sec:drl_results}
\vspace*{0.25cm}

\begin{figure}[!t]
    \centering
    \begin{subfigure}[t]{.45\linewidth}
        \centering
        \includegraphics[width=\linewidth]{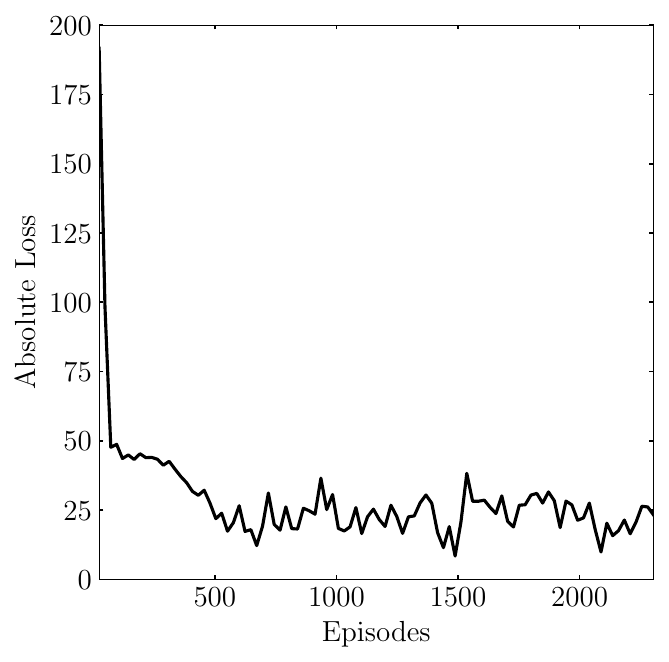}
    \end{subfigure}
    \begin{subfigure}[t]{.45\linewidth}
        \centering
        \includegraphics[width=\linewidth]{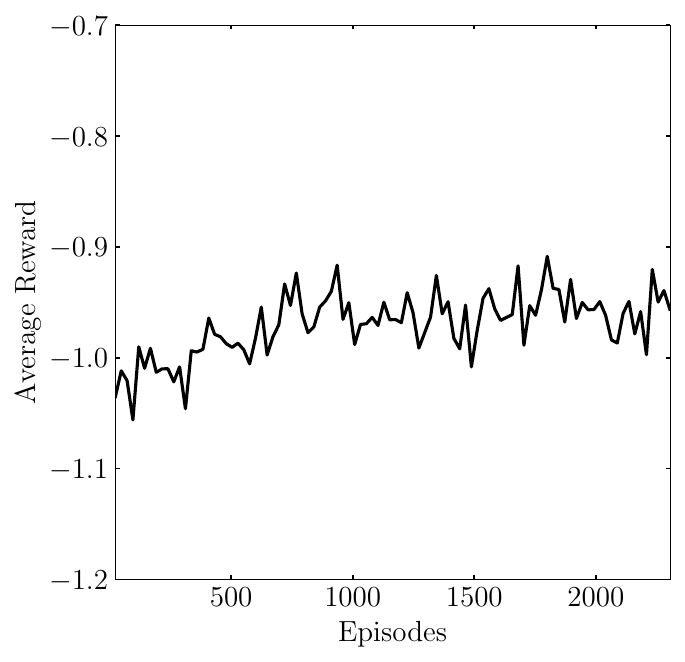}
    \end{subfigure}
    \begin{subfigure}[t]{.45\linewidth}
        \centering
        \includegraphics[width=\linewidth]{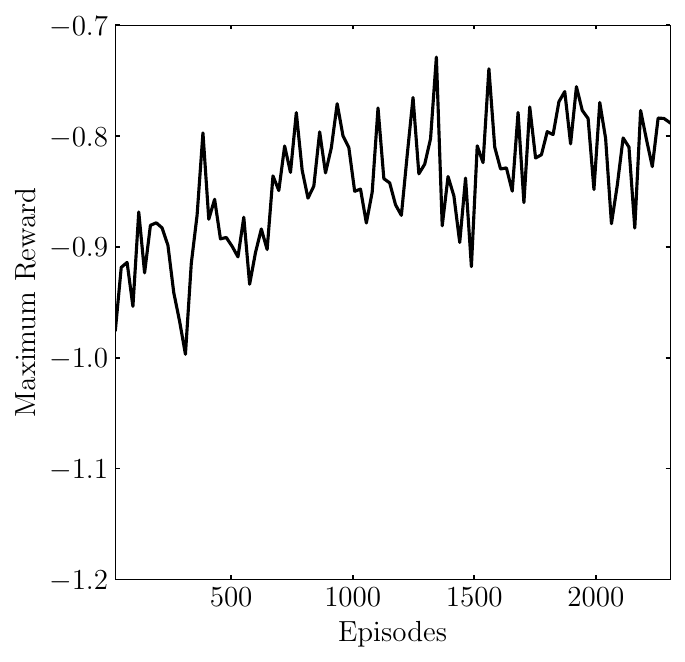}
    \end{subfigure}
    \begin{subfigure}[t]{.45\linewidth}
        \centering
        \includegraphics[width=\linewidth]{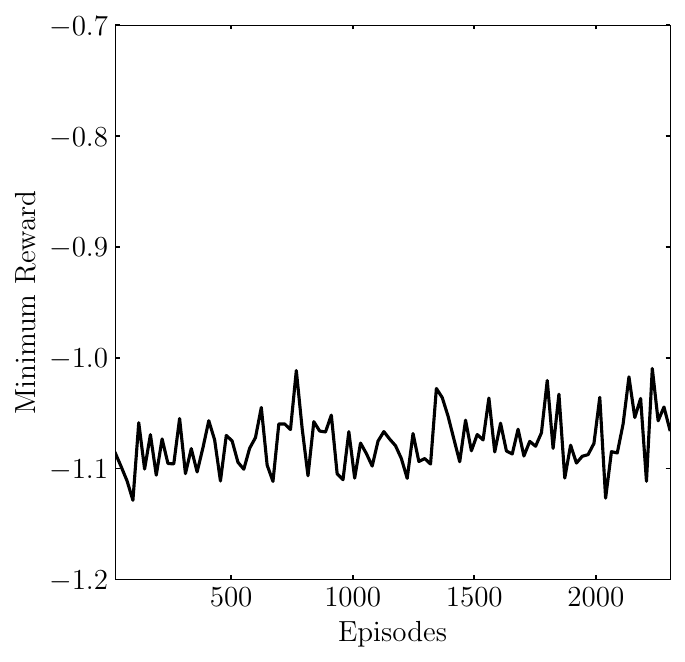}
    \end{subfigure}
    \caption{DRL training metrics. The average, maximum, and minimum reward signals are extracted from the 24 pseudo-environments running in parallel.}
    \label{fig:drl_training}
\end{figure}

In the following, the results of the DRL control are discussed. First, the training of the DRL model is shown in figure \ref{fig:drl_training}. It can be observed that the model's loss correctly decreases with increasing number of episodes before it stagnates. The average reward metric is the average value of the time-averaged characteristic recirculation lengths provided by the pseudo-environments after an episode. It is observed that the average reward improves with training (bearing in mind that, during training, exploration noise is added to the controlling agent) and converges to an average reward lower than $-1$, hence reducing the TSB compared to the non-actuated baseline case. An improvement is also observed on the maximum reward, while the minimum reward remains stable. As previously explained, 24 trajectories are sampled in parallel since $N_\mathrm{e}=8$ and $N_\mathrm{pe}=3$, and an equal number of optimization steps (24) are performed thereafter. This process is repeated 96 times, hence accumulating a total of 2304 episodes and optimization steps. Noting that the training time for a single CFD environment takes 1.5 hours to simulate 1600 TU on a A100 NVIDIA GPU, the overall training walltime is 144 hours (6 days) when running 8 CFD environments in parallel (one per GPU). This results in a total of 1152 GPU-hours consumed.

\begin{figure}[!t]
    \centering
    \begin{subfigure}[t]{.32\linewidth}
        \centering
        \includegraphics[width=\linewidth]{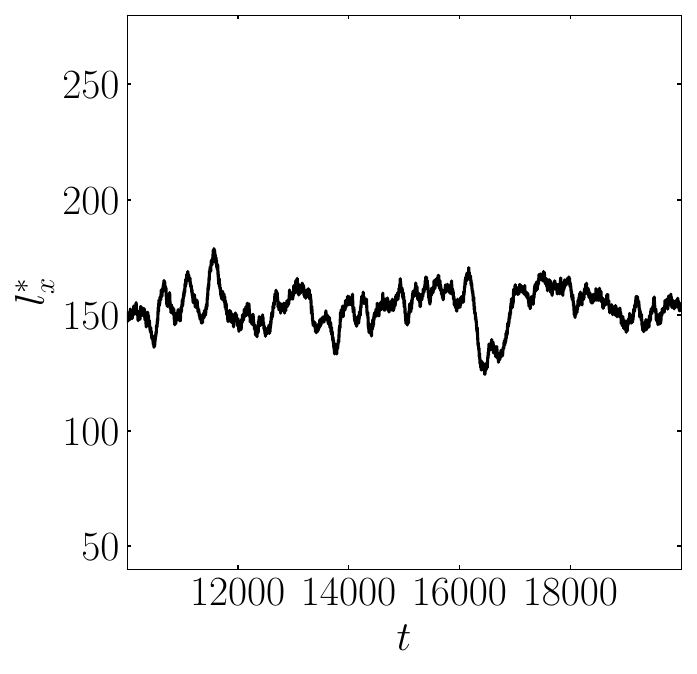}
    \end{subfigure}
    \begin{subfigure}[t]{.32\linewidth}
        \centering
        \includegraphics[width=\linewidth]{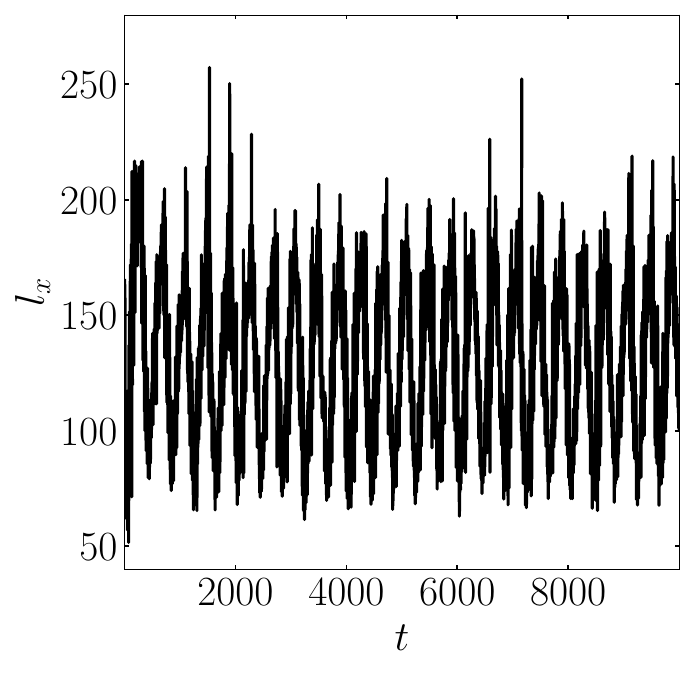}
    \end{subfigure}
    \begin{subfigure}[t]{.32\linewidth}
        \centering
        \includegraphics[width=\linewidth]{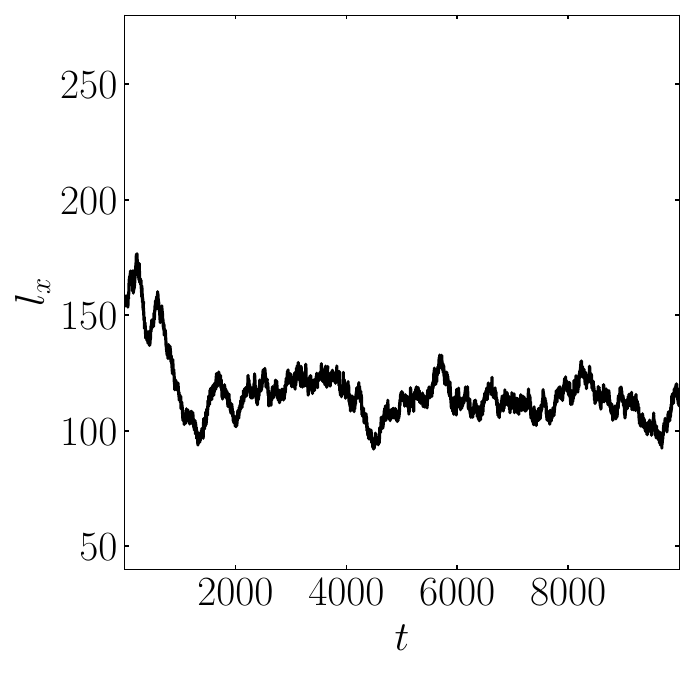}
    \end{subfigure}
    \caption{Temporal signal of the characteristic recirculation length for the non-actuated (left), periodic control (center), and the DRL control (right) coarse-grid cases. The periodic control and DRL control cases use the non-actuated snapshot at $t=20000$ as initial condition.}
    \label{fig:lx_drl}
\end{figure}

Once the DRL agent has been trained, we test the learnt control strategy under deterministic behaviour, {\it i.e.} selecting the best possible action from the action probability distribution (its mean). The DRL model is run in deterministic mode for 20000 TU using a baseline (non-actuated) snapshot as initial condition. Table \ref{tab:lx_drl} shows the average characteristic recirculation length of the non-actuated, periodic control, and DRL control cases for the last 15000 TU. The DRL control yields a larger reduction of the TSB compared to the classic periodic control, respectively $-25.3\%$ and $-15.7\%$. The temporal signal $l_x$ is plotted in figure \ref{fig:lx_drl} for the first 10000 TU. It is clearly appreciated that the DRL control quickly reduces the TSB length. Importantly, the control is performed smoothly, and no sudden oscillations are appreciated. While the periodic forcing also displays a reduction of the bubble length, spurious peaks can be observed in the signal (this is better appreciated in figure \ref{fig:lx}). This can arise from the fact the DRL control strategy is forced to conserve mass in space, {\it i.e.} instantaneously and therefore in time too, while the classical periodic forcing is only conserving mass in time. The instantaneous conservation of mass in incompressible flow is important not only from the physical point of view, but also numerically in the pressure solver. Since we do not correct for the instantaneous mass imbalance of the classical periodic control, spurious oscillations can arise, which are then captured by the $l_x$ signal.

\begin{figure}[t!]
    \centering
    \begin{subfigure}[t]{\linewidth}
        \centering
        \includegraphics[width=\linewidth]{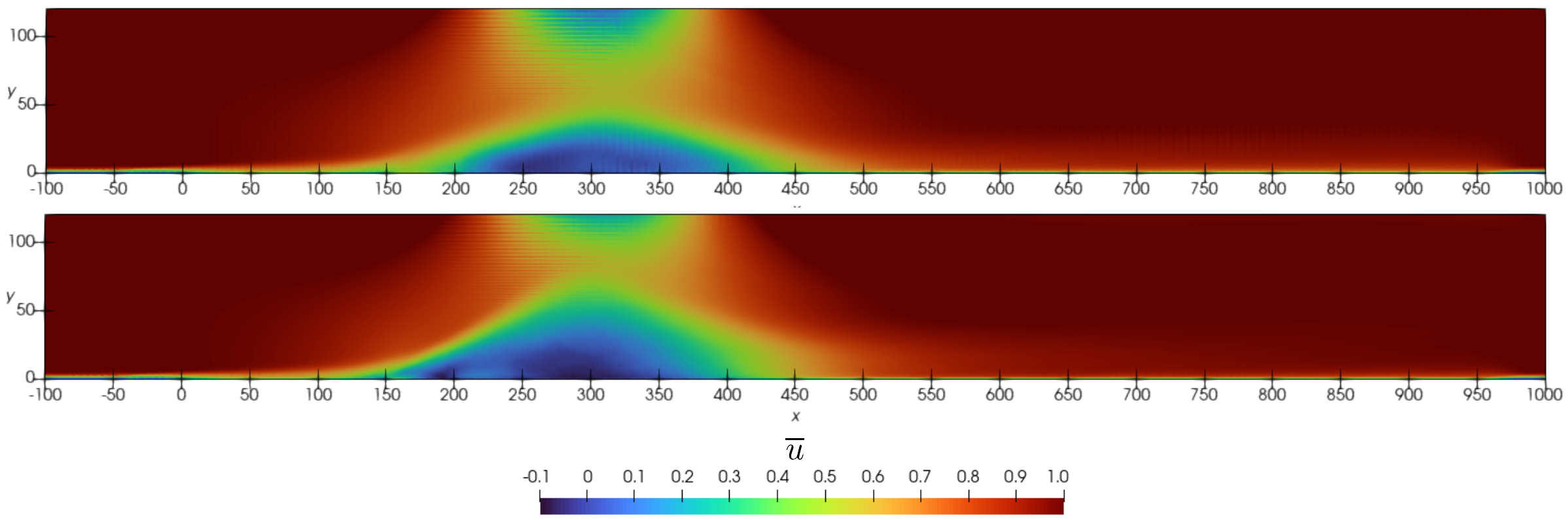}
        \caption{Time-averaged streamwise velocity component $\overline{u}$ at $z=L_z/2$.}
        \label{fig:front_coarse_znmf_drl}
    \end{subfigure}
    \begin{subfigure}[t]{\linewidth}
        \vspace*{0.2cm}
        \centering
        \includegraphics[width=\linewidth]{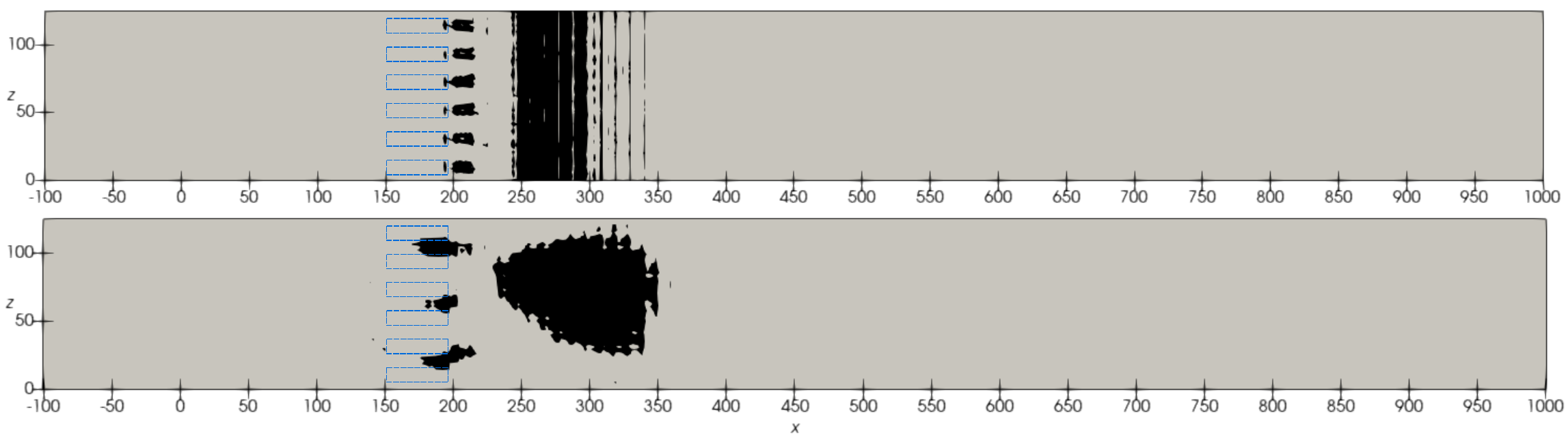}
        \caption{Time-averaged recirculation region ($\overline{u}<0$) at the $xz$-plane of the first off-wall node.}
        \label{fig:bubble_coarse_znmf_drl}
    \end{subfigure}
    \caption{Coarse-grid time-averaged flow field. Both panels (a) and (b) show periodic control (top) and DRL control (bottom) results (periodic control results are shown again for better comparison).}
\end{figure}

Flow fields for both periodic control and DRL control are shown in figure \ref{fig:front_coarse_znmf_drl} (results of the periodic control are shown again for comparison purposes). The DRL control yields a different flow-field structure. A strong recirculation region located at $x=180$, on top of the actuators, is observed. Since DRL actuators are spanwise-paired with equally positive and negative mass flow rates, this can eventually generate streamwise structures that interact with the TBL resulting in these small separation bubbles. This phenomenon is confirmed in figure \ref{fig:bubble_coarse_znmf_drl}, and it can also be noted that the bubble reduction of the DRL control is qualitatively more prominent than the periodic-forcing control.

\begin{figure}[h]
    \centering
    \includegraphics[width=\linewidth]{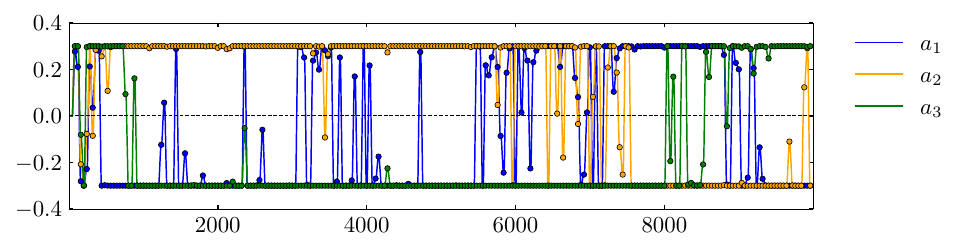}
    \caption{Temporal signal of the DRL-based actuators. The exponential smoothing between two discrete actions that sets the instantaneous actuators mass flow rate is also displayed.}
    \label{fig:actions}
\end{figure}

Last, the time signals of the actions set by the DRL control agent are shown in figure \ref{fig:actions}. First, the actuators oscillate between the maximum and minimum allowed values, $A_\mathrm{ac}=\pm0.3$. Shorty after, the $a_3$ actuator saturates on $+A_\mathrm{ac}$ and $a_2$ follows, while $a_1$ saturates on $-A_\mathrm{ac}$. Then, actuators bounce between $+A_\mathrm{ac}$ and $-A_\mathrm{ac}$, {\it i.e.} bang--bang control, while short transitional phases arise as well. A physical interpretation of the DRL actions cannot be derived from the current results yet, and a thorough assessment of the control strategy learnt by the DRL agent will be considered in future work.

\begin{table}[t!]
    \begin{center}
        \begin{tabular}{llc}
        \br
         &\thead{$\overline{l_x}$} & $1-\overline{l_x}/\overline{l_x^*}$\\
        \mr
        Periodic & $129.5\pm33.4$ & 15.7\%\\
        DRL & $114.7\pm6.7$ & 25.3\%\\
        \br
        \end{tabular}
    \end{center}
    \caption{Time-averaged characteristic length $\overline{l_x}$ and standard deviation of periodic control and DRL control cases. The reduction of the recirculation length wrt. the non-actuated case is also shown.}
    \label{tab:lx_drl}
\end{table}

\newpage
\section{Conclusions}
\vspace*{0.25cm}
\label{sec:conclusions}
In the present work, we investigate the efficacy of classical periodic control and deep reinforcement learning (DRL) control in reducing a turbulent separation bubble (TSB) generated by a suction and blowing (SB) boundary condition. The framework that integrates the SOD2D CFD solver (a novel multi-GPU spectral element solver developed at Barcelona Supercomputing Center) with the DRL model via SmartSim is also presented.

The classical periodic control, employing harmonic time forcing, provides notable success in reducing the TSB length by approximately $6.8\%$ and $15.7\%$ on a fine and a coarse grid, respectively. Additionally, a spectral analysis highlights the impact of the actuators on the flow, with clear peaks at the actuation frequency and harmonics. The coarse-grid results demonstrate the preservation of the main flow features, validating the use of the coarse grid for the subsequent training of the DRL model. While the current periodic forcing considers a single configuration of frequency and amplitude, future work will expand this parametric space in order to maximize the periodic control efficacy. An investigation of the domain size shall also be explored to assess the confinement of the bubble for both non-actuated and actuated cases.

The DRL control demonstrated very promising results by successfully learning a control strategy that can effectively reduce the TSB length by $25.3\%$ in the coarse grid. Compared with the classical periodic control, the DRL has the freedom to set the optimal mass flow rate of the actuators for a given environment state, hence constructing a complex control signal that can embed multiple frequencies. The training was performed using 24 parallel pseudo-environments running on 8 GPUs and took 144 hours (6 days) in total, so 1152 GPU-hours. The physical interpretation of the control strategy remains for future work, as well as the evaluation of the DRL strategy on the fine mesh. Nonetheless, to the best of our knowledge, the current results present a successful application of DRL-based control at one of the highest Reynolds numbers to this date.

\section{Software}
The software used in this work is open-source, and the main packages that compose our CFD--DRL framework are listed below.
\begin{itemize}
    \item \textbf{SOD2D}: the multi-GPU CFD solver based on the spectral element method. Available in \url{https://gitlab.com/bsc_sod2d/sod2d_gitlab}.
    \item \textbf{SmartSOD2D}: the communications package for SOD2D based on SmartSim. It allows online training of ML models among other co-processing possibilities. Available in \url{https://github.com/b-fg/SmartSOD2D}.
    \item \textbf{SmartSim}: the workflow library to deploy ML on HPC applications. Available in \url{https://github.com/CrayLabs/SmartSim}.
    \item \textbf{TF-Agents}: the reinforcement learning library based on TensorFlow. Available in \url{https://github.com/tensorflow/agents}.
\end{itemize}

\ack
\vspace*{0.25cm}
This work was performer in part during the Fifth Madrid Summer Workshop, funded by the European Research Council (ERC) under the Caust grant ERC-AdG-101018287. OL has been partially funded by the European Commission's Horizon 2020 Framework program and the European High-Performance Computing Joint Undertaking (JU) under grant agreement no. 101093393, and by MCIN/AEI/10.13039/501100011033 and the European Union NextGenerationEU/PRTR (PCI2022-134996-2), project CEEC. OL has been partially supported by a Ramon y Cajal postdoctoral contract (Ref: RYC2018-025949-I). RV and FAA acknowledge financial support from ERC grant no. `2021-CoG-101043998, DEEPCONTROL'. The authors thank Laia Juli\'o from the BSC support services and Dr. Andrew Shao from the SmartSim team for the help and fruitful discussions about the CFD--DRL software framework. The authors also thank the UPM Turbulence Summer School and our hosts Dr. Miguel P\'erez Encinar and Dr. Adrián Lozano-Durán for the organization of the workshop. Finally, we acknowledge the National Academic Infrastructure for Supercomputing in Sweden (NAISS) for the computational time provided at the large-GPU systems Alvis and Berzelius.

\section*{References}
\vspace*{0.25cm}
\bibliography{main}

\end{document}